\documentclass[epj, twocolumn]{webofc}
\usepackage[varg]{txfonts}
\pdfoutput=1

\woctitle{19th Joint Workshop on Electron Cyclotron Emission and Electron Cyclotron Resonance Heating}

\begin{document}

\title{A Kinetic Study of Microwave Start-up of Tokamak Plasmas}

\author{\firstname{E.J.} \lastname{du Toit}\inst{1,2}\fnsep\thanks{\email{ejdt500@york.ac.uk}}
 \and
        \firstname{M.R.} \lastname{O'Brien}\inst{2}
 \and
        \firstname{R.G.L.} \lastname{Vann}\inst{1}
}

\institute{York Plasma Institute, Department of Physics, University of York, York, YO10 5DD, UK 
\and
           Culham Centre for Fusion Energy, Abingdon, OX14 3DB, UK
}

\abstract{%
  A kinetic model for studying the time evolution of the distribution function for microwave start-up is presented. The model for the distribution function is two dimensional in momentum space, but, for simplicity and rapid calculations, has no spatial dependence. Experiments on the Mega Amp Spherical Tokamak have shown that the plasma current is carried mainly by electrons with energies greater than $70$ keV, and effects thought to be important in these experiments are included, i.e. particle sources, orbital losses, the loop voltage and microwave heating, with suitable volume averaging where necessary to give terms independent of spatial dimensions. The model predicts current carried by electrons with the same energies as inferred from the experiments, though the current drive efficiency is smaller.
}

\maketitle

\section{Introduction}
Non-inductive plasma current start-up is a very important area of research for the spherical tokamak (ST) due to a lack of space for a shielded inboard solenoid. Reliable models are needed to predict performance and start-up requirements for present and future STs. This paper describes development of such a model for electron Bernstein waves, and first comparison of its predictions with experimental results.

\begin{figure}[!hbt]
	\centering
	\includegraphics[width=0.45\textwidth]{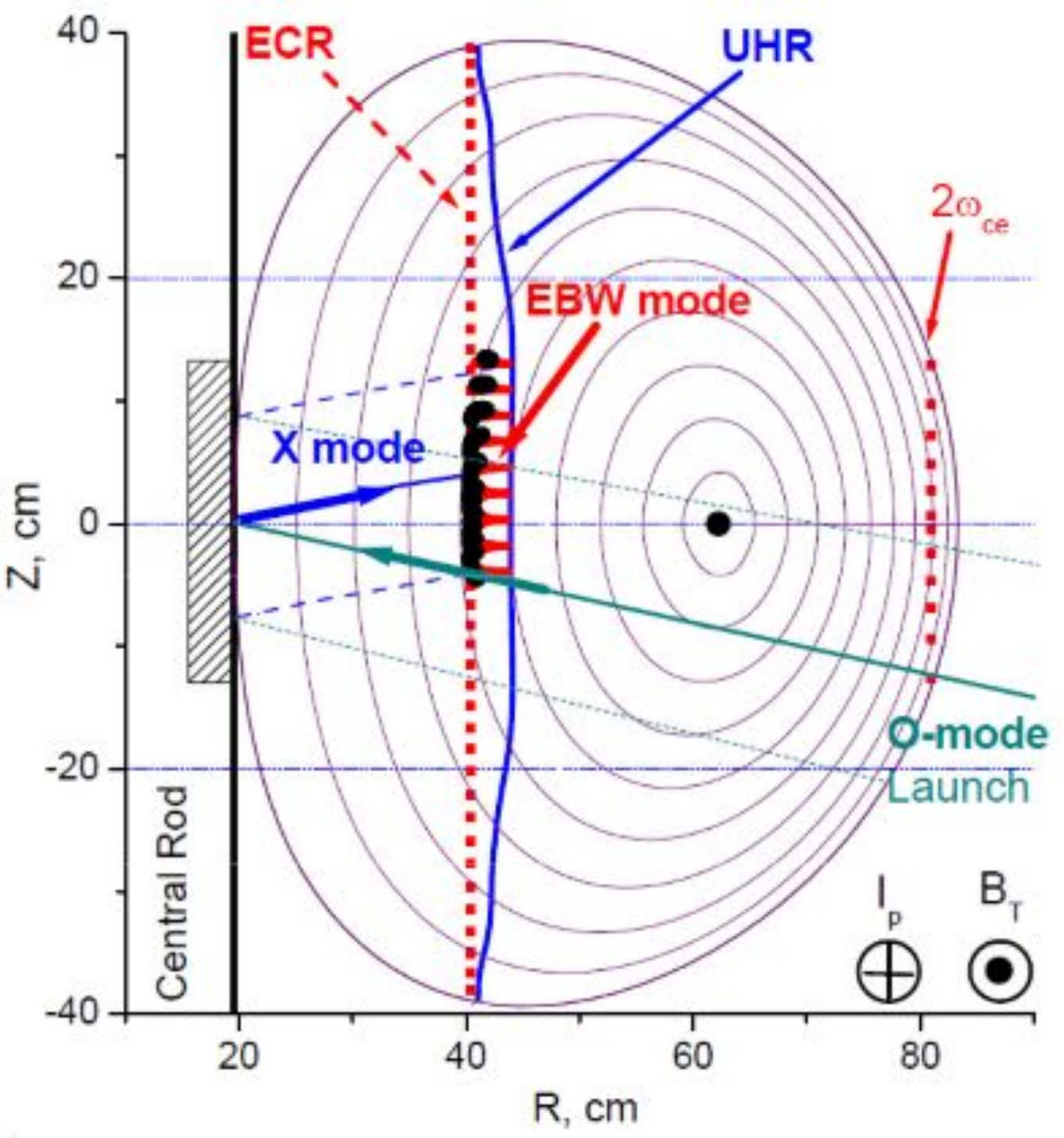}
	\caption[]{Schematic of the EBW assisted plasma current start-up. The black area at the end of the EBW rays indicates the power deposition zone \cite{Shevchenko_2010}.}
	\label{fig:MAST}
\end{figure}

Electron Bernstein wave (EBW) start-up using a \\ $28$ GHz gyrotron capable of delivering $100$ kW for up to $0.5$ s was demonstrated on the Mega Amp Spherical Tokamak (MAST) \cite{Shevchenko_2010, Shevchenko_2015}. The EBW start-up method employed here relied on the production of low-density plasma by RF pre-ionization around the fundamental electron cyclotron resonance (ECR). The toroidal magnetic field generated by the central rod (CR), generated by a current of about $2$ MA, had a value of $B_\phi = 1$ T at a radial position of about $0.4$ m. A double mode conversion (MC) was used for EBW excitation. The scheme consisted of MC of the ordinary (O) mode, incident from the low field side of the tokamak, into the extraordinary (X) mode with the help of a grooved mirror-polarizer incorporated in a graphite tile on the CR. The launched Gaussian beam was tilted to the midplane at $10^\circ$ and hit the CR at the midplane, as illustrated in figure \ref{fig:MAST}. The X-mode reflected from the CR propagated back to the plasma, passed through the ECR and experienced a subsequent slow X to EBW MC near the upper hybrid resonance (UHR). The excited EBW mode was totally absorbed before it reached the ECR, due to the Doppler shifted resonance. Modelling showed that only a small fraction of the injected power $(\sim2\%)$ was absorbed as the O- and X-modes, while the main part was converted into and then absorbed as the EBW mode \cite{Shevchenko_2015}. 

The various experiments conducted on EBW assisted start-up are well documented in \cite{Shevchenko_2010} and \cite{Shevchenko_2015}, as well as the references therein. The main conclusions were: firstly, the formation of closed flux surfaces (CFS) is governed by EBW current drive (CD); secondly, the plasma current is carried predominantly by supra thermal electrons with energies above $70$ keV; and thirdly, there exists a linear relationship between the injected RF power and the generated plasma current, such that an overall current drive efficiency of $1$ A/W was achieved.

Models are important to interpret and predict start-up current drive in STs, and confirm conclusions drawn from experiments. To this end, a start-up model has been developed to study the time evolution of the electron distribution function. For simplicity, and to allow for rapid calculations, the distribution function has zero spatial dependence, while it is two dimensional in momentum space. The effects thought to be important in capturing the main physics during the early stages of the plasma discharge have been included, i.e. particle sources, orbital losses, the loop voltage and EBW heating, with suitable volume averaging where necessary to give terms independent of spatial dimensions.

The paper is structured as follow: an introduction to the kinetic model describing the time evolution of the electron distribution function for studying EBW start-up is presented, with special attention being given to particle losses and EBW heating, and how the spatial dependences are approximated in $0$D in these terms. This is followed by a discussion of results and its comparison to experiments done on MAST, and a brief overview of future work.

\section{0D2V kinetic start-up model}
In general, the electron distribution function depends on space, momentum, and time, $f\left( \vec{r},\vec{p},t \right)$. The distribution function can be used to calculate the current density,
	\begin{equation*}
	J\left( \vec{r},\vec{t} \right) = e \int \textrm{d}^3 p \, v_\parallel \, f\left( \vec{r},\vec{p}, t \right)
	\end{equation*}
and the electron density,
	\begin{equation*}
	n_e = \int \textrm{d}^3 p \, f\left( \vec{r},\vec{p}, t \right).
	\end{equation*}

In order to study the time evolution of the distribution function, simplifications are made to ensure the model is tractable and computationally manageable. For this reason, the model is zero dimensional ($0$D) in space, but spatial dependences for each term are taken into account when calculating the $0$D approximations. Further, the time it takes for an electron to complete a gyro-orbit is fast compared to all other timescales, such that the momentum dependence can be captured in two dimensions ($2$V). The time evolution of the electron distribution function is then studied under several effects,
	\begin{equation}
	\frac{\partial f}{\partial t} = \textrm{source} + \textrm{loss} + \textrm{EBW heating} + \textrm{loop voltage}
	\end{equation}
where $f = f\left( p_\parallel, p_\perp, t \right)$.

Experiments suggest that the plasma current is carried by supra thermal electrons with energies in the range $70-100$ keV. Under typical MAST parameters, these electrons collide very infrequently (for electrons with energies above $20$ keV, the collision times are longer than $100$ ms), and therefore collisions are neglected in this paper, while the terms expected to be most important in studying the time evolution of the distribution function are retained.

\begin{figure}[!hbt]
	\centering
	\includegraphics[width=0.45\textwidth]{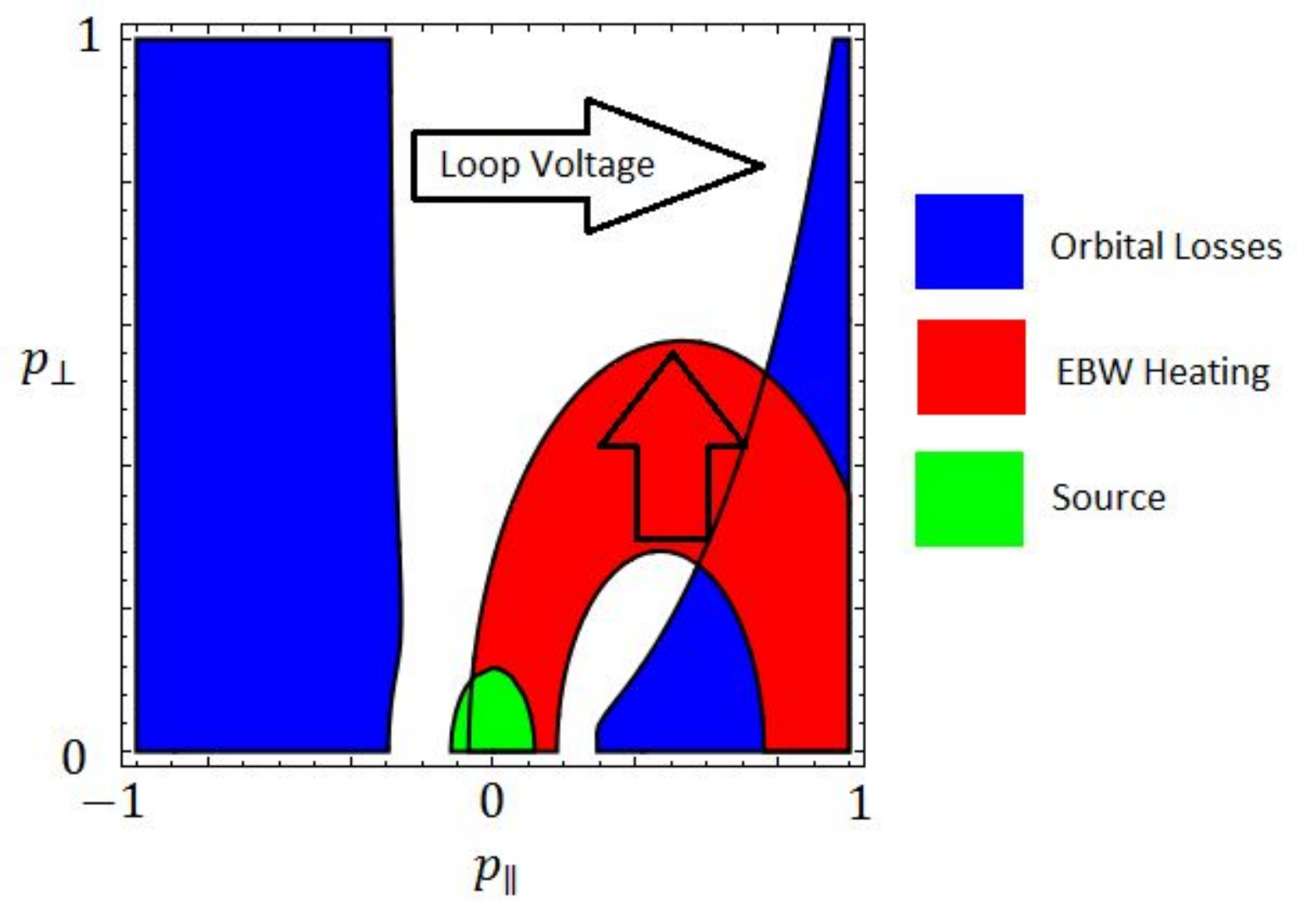}
	\caption[]{Schematic of the effects under which the time evolution of the distribution function is studied.}
	\label{fig:model}
\end{figure}

Each term acts in a different region of momentum space, as illustrated in figure \ref{fig:model}. The source term is due to cold electrons entering the system, resulting mainly from ionization. These electrons are taken to be isotropic in momentum, such that
	\begin{equation*}
	\left( \frac{\partial f}{\partial t} \right)_\textrm{source} = \frac{S_{\!0}}{\pi^{3/2} p_0^3} \exp{\left[ - \frac{p_\parallel^2 + p_\perp^2}{p_0^2} \right]}
	\end{equation*}
where $p_0$ is the characteristic momentum of these cold electrons, and $S_{\!0}$ is the rate of change in electron density.

The loss term is due to electrons being lost out of the plasma volume, described in section \ref{sec:loss}. The EBW heating term describes the interaction between the plasma and the injected RF beam, described in section \ref{sec:RF}.

The electric field created by the possible use of the central solenoid results in an acceleration of electrons to higher parallel momentum, and is described by the loop voltage term, given by
	\begin{equation*}
	\left( \frac{\partial f}{\partial t} \right)_\textrm{loop voltage} = -e \frac{V_L}{2 \pi R} \frac{\partial f}{\partial p_\parallel}
	\end{equation*}
where $V_L$ is the magnitude of the loop voltage, and $R$ is the major radius.

\subsection{Orbital Losses}
\label{sec:loss}
An important part of the start-up scenario is the transition from an open field line configuration to closed flux surfaces (CFS). During the initial start-up phase, the magnetic field lines are considered to be open, and electrons can freely stream along these field lines out of the plasma volume into the vessel walls. This loss is due to the parallel velocity of the electrons, and depends on the vertical strength of the magnetic field, such that a characteristic confinement time for an electron with parallel velocity $v_\parallel$ can be modelled as \cite{Kim_2012},
	\begin{equation*}
	\tau_\parallel = a_\perp \frac{B_\phi}{B_Z} / v_\parallel
	\end{equation*}
where $a_\perp$ is the perpendicular distance to the vessel walls. This term has to be modified to allow for the formation of CFS, as particle confinement drastically improves when CFS form. An additional factor is therefore added,
	\begin{equation}
	\tau_\parallel = a_\perp \exp{\left( \frac{I_p}{I_\textrm{ref}} \right)} \frac{B_\phi}{B_Z} / v_\parallel
	\end{equation}
where $I_p$ is the plasma current and $I_\textrm{ref}$ is the magnitude of the plasma current where CFS form \cite{Kim_2012}.

After CFS form, the loss mechanism changes from mainly parallel transport to mainly perpendicular transport. Collisions can transport electrons across CFS, such that some electrons are lost. Bohm diffusion is used to describe perpendicular transport \cite{Wauters_2011}, such that a characteristic perpendicular loss time is given by,
	\begin{equation}
	\tau_\perp = \frac{8 a^2}{T_e(eV)} B_\phi
	\end{equation}
where $a$ is the minor radius of the plasma and $T_e$ is the energy of the electron.

\begin{figure}[!hbt]
	\centering
	\includegraphics[width=0.45\textwidth]{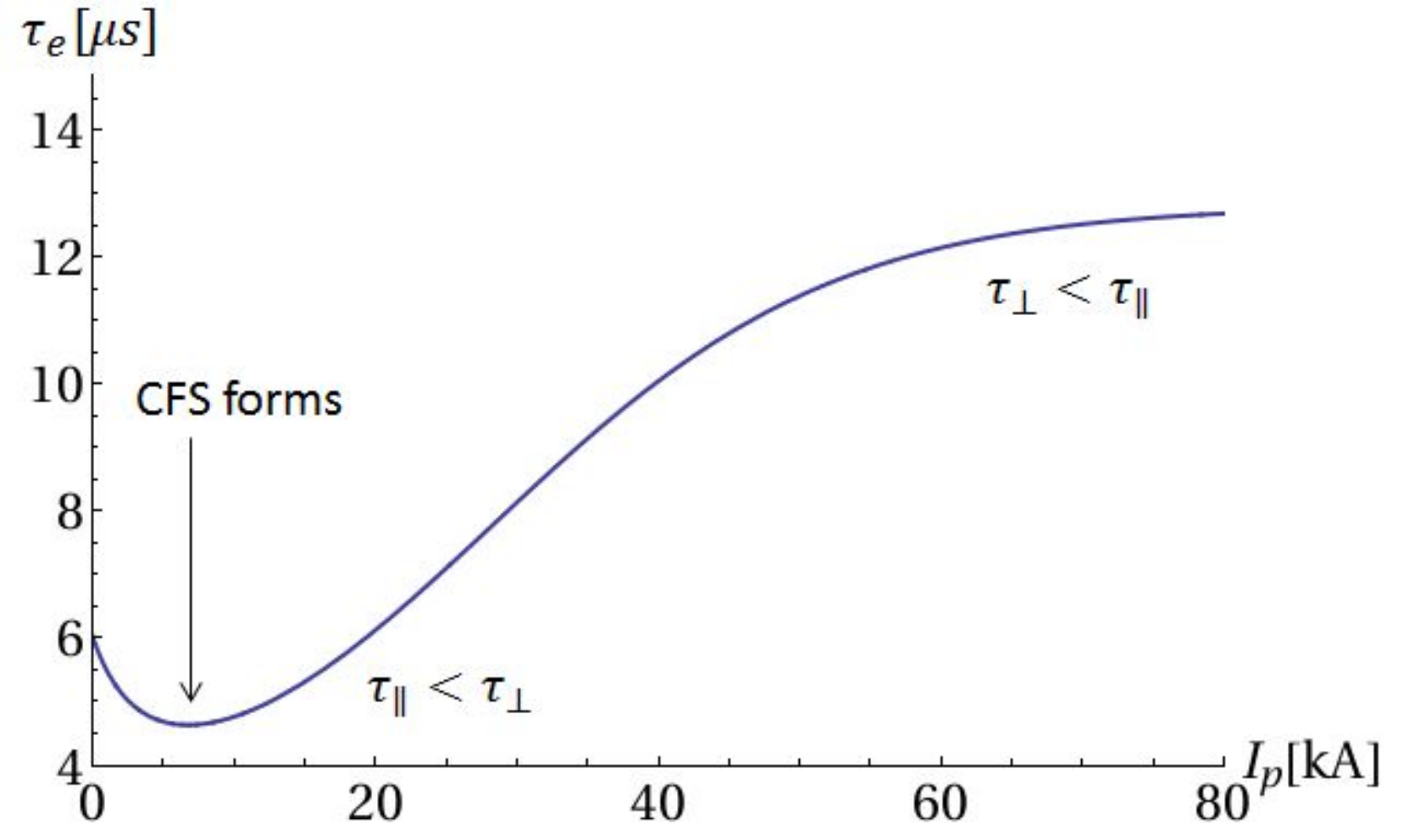}
	\caption[]{Characteristic particle loss time for an energetic electron ($E_k = 50$ keV) during start-up phase as a function of plasma current. At small currents, before closed flux surfaces form, parallel transport dominates, while perpendicular transport dominates for large currents.}
	\label{fig:lossTime}
\end{figure}

The confinement time of an electron is therefore given by,
	\begin{equation}
	\label{eq:lossTime}
	\frac{1}{\tau_e} = \frac{1}{\tau_\parallel} + \frac{1}{\tau_\perp}
	\end{equation}
with the result for an electron of a specific energy shown in figure \ref{fig:lossTime}. 

This loss time, however, does not take into account the magnetic field line structure leading to effects such as trapped electrons. It can therefore only be used as an estimation of the particle loss time, as a function of plasma current, while the momentum dependence has to be added explicitly. For this reason, the distribution function is subjected to a loss term of the form,
	\begin{equation}
	\left( \frac{\partial f}{\partial t} \right)_\textrm{loss} = - \frac{f}{\tau_e} A(p_\parallel,p_\perp)
	\end{equation}
where $A(p_\parallel,p_\perp)$ is the functional form of the loss term, containing the momentum dependence, while the loss time $\tau_e$ determines the magnitude of the loss term.

\begin{figure}[!hbt]
	\centering
	\includegraphics[width=0.45\textwidth]{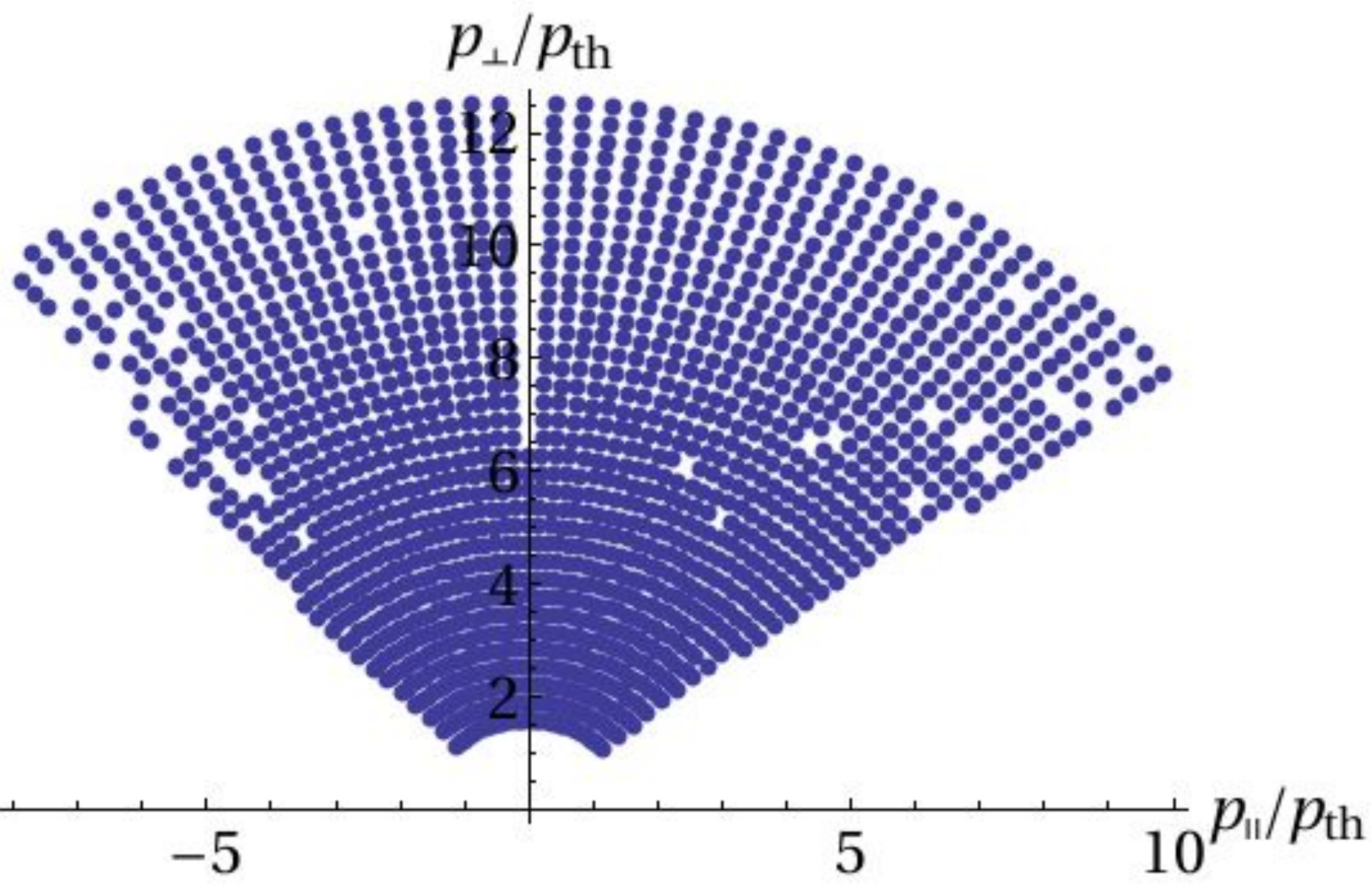}
	\caption[]{Confinement map for electrons originating from the same point on the poloidal plane show a slight asymmetry. Electrons initially moving against the magnetic field are lost more easily than electrons moving along the magnetic field.}
	\label{fig:lossForm}
\end{figure}

The momentum dependence of the loss term is found by studying single particle orbits during an open field line structure. By studying the guiding centre orbits of electrons originating from the same point in space, the time it takes for these electrons to be lost or to complete a confined orbit can be calculated, and plotted as a function of momentum. The time it takes for electrons to be lost out of the plasma volume, calculated by the guiding centre approach, is consistent with the time found from equation (\ref{eq:lossTime}). Figure \ref{fig:lossForm} shows the initial values of momentum for which electrons complete a confined orbit, while all other electrons will be lost out of the plasma volume. An approximate form for the loss term can then be determined,
	\begin{equation}
	A(p_\parallel, p_\perp) = \frac{p_\parallel^2 + p_\perp^2}{p_\textrm{loss}^2} \left\{ \, \begin{matrix}  \left( 1 - \exp{ \left[ - \frac{5 p_\parallel^2}{4 p_\perp^2} \right] } \right) & \quad  p_\parallel < 0 \\
													 \left( 1 - \exp{ \left[ - \frac{4 p_\parallel^2}{5 p_\perp^2} \right] } \right) & \quad  p_\parallel \ge 0
					\end{matrix} \right.
	\end{equation}
where $p_\textrm{loss}$ is the characteristic momentum of an electron with loss time $\tau_e$.

The typical confinement map for electrons show that the confinement time decreases for increasing parallel momentum, while confinement improves when increasing the perpendicular momentum. There is also an asymmetry in the loss term, for electrons moving with and against the magnetic field, which will disappear once CFS form.

\subsection{EBW Heating}
\label{sec:RF}
The interaction between the injected RF beam and the plasma is described by the EBW heating term. This form of heating mainly increases the perpendicular momentum of electrons,
	\begin{equation}
	\left( \frac{\partial f}{\partial t} \right)_\textrm{EBW heating} = \frac{1}{p_\perp} \frac{\partial}{\partial p_\perp} p_\perp \, D(p_\parallel,p_\perp) \, \frac{\partial f}{\partial p_\perp}
	\end{equation}
where the diffusion coefficient is assumed to take the form,
	\begin{equation}
	D(p_\parallel,p_\perp) = D_0 \left\langle \exp{\left[- \left( \frac{p_\parallel - p_{\parallel 0}}{\Delta p_\parallel} \right)^2 \right]} \right\rangle_\textrm{volume average}
	\end{equation}
and $D_0$ is a constant to be determined. It is assumed that the EBW heating is localised in momentum space, and that it occurs where the resonance condition $(\omega - k_\parallel v_\parallel - n \omega_c = 0)$ is satisfied. For a particular value of $p_\perp$, $p_{\parallel 0}$ is the parallel momentum that satisfies the resonance condition. The absorption width $\Delta p_\parallel$ is due to a spread in $k_\parallel$, the parallel component of the wave number of the EBW.

\begin{figure}[!hbt]
	\centering
	\includegraphics[width=0.45\textwidth]{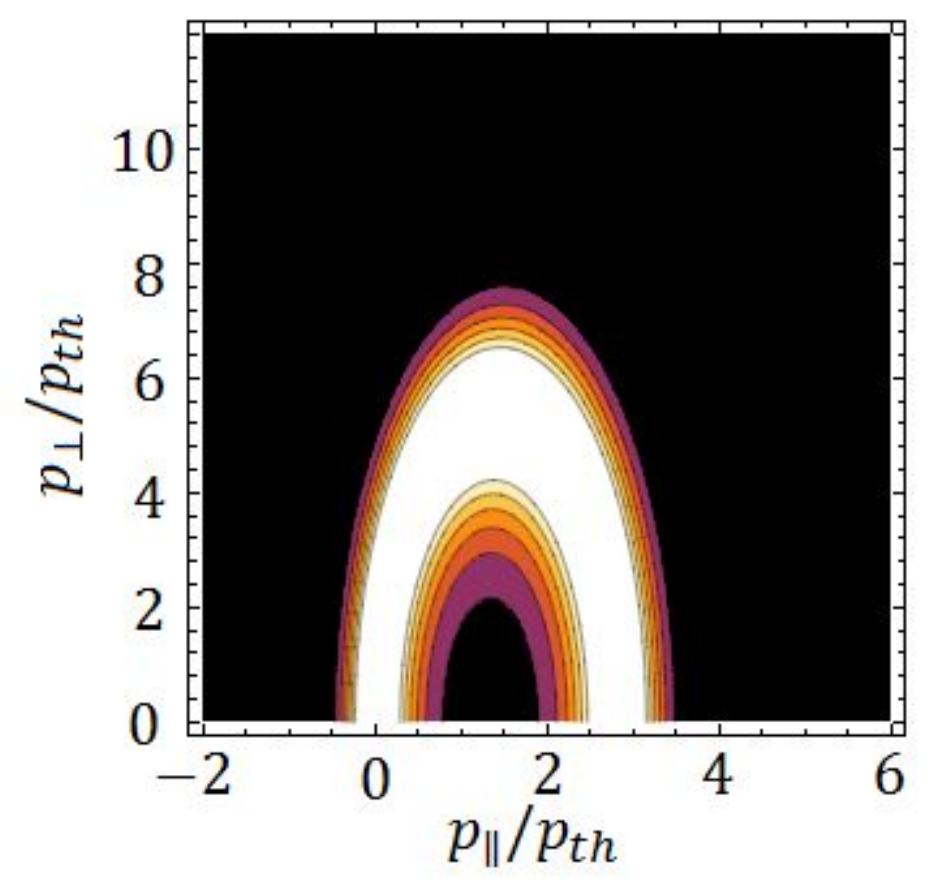}
	\caption[]{Typical resonance curve for EBW heating shows the location where the plasma wave interaction is prominent. The diffusion term is smeared out in momentum space due to the spatial dependence of the resonance condition, while a maximum value of $p_\perp$ beyond which the resonance condition can no longer be satisfied is clearly visible.}
	\label{fig:resonance}
\end{figure}

EBW heating is not uniform in space, due to the dependence of the relativistic cyclotron frequency \\ $\omega_c = q_e B/m_e \gamma$ on the magnetic field. The spatial dependence of the plasma wave interaction has to be resolved, and this is done by integrating over the plasma in taking a volume average,
	\begin{equation*}
		\begin{aligned}
	\left\langle \exp{\left[- \left( \frac{p_\parallel - p_{\parallel 0}}{\Delta p_\parallel} \right)^2 \right]} \right\rangle_\textrm{volume average} &= \\
	\frac{1}{V} \int_{\textrm{plasma}} \textrm{d}V &\exp{\left[- \left( \frac{p_\parallel - p_{\parallel 0}}{\Delta p_\parallel} \right)^2 \right]}
		\end{aligned}
	\end{equation*}
such that the plasma wave interaction is limited to the region in space where the plasma exists. This produces a momentum dependent diffusion term $D(p_\parallel,p_\perp)$ independent of space, illustrated in figure \ref{fig:resonance}. The diffusion term is largest along the resonance curve (the solution to the resonance condition) where absorption will be maximal, but smeared out in momentum space due to the spatial dependence of the resonance condition, as the resonance curve is slightly different for different values of the magnetic field.

The power absorbed by the plasma from the EBW can be calculated from the distribution function,
	\begin{equation}
	P_d = \frac{1}{2} m_e \int \textrm{d}V \int \textrm{d}^3 p \left( \frac{\partial f}{\partial t} \right)_\textrm{EBW heating}
	\end{equation}
where the volume integral is introduced to obtain the total power absorbed, rather than the power density. This can be related to the absorption efficiency of the plasma through the optical depth $\sigma$:
	\begin{equation}
	\label{eq:EBW:balance}
	\sigma P_0 = P_d
	\end{equation}
where $P_0$ is the injected RF power. In general, radiation can be reflected in the tokamak to give multi-pass absorption, but as the plasma volume is assumed to be small compared to the vessel volume during start-up, this effect is assumed to be small, and absorption is approximated as single pass.

For ECRH, standard expressions for $\sigma$ exist in the literature (see, for example \cite{Bornatici_1983}), such that the absorption efficiency $\sigma P_0$ can readily be calculated. For EBW, absorption is expected to be much stronger, such that $\sigma$ would be approximately $1$. The value of $\sigma$, in this case, would then be related to the amount of power converted to EBW, such that the expression (\ref{eq:EBW:balance}) remains the same. The value of $D_0$ can then be chosen in order to satisfy this equation, by ensuring the distribution function produces the correct value of $P_d$.

This method can be used to study both ECRH and EBW start-up. The wave parameters in these two cases of heating are different, but the diffusion coefficient is related to the resonance condition, which depends on the particular wave parameters, while the absorption is related to the optical depth in the case of ECRH, or to the amount of power converted to EBW. This is a rather versatile method in that sense, as the terms and their forms do not change depending on the type of heating, but only the values.

\section{EBW Start-up}
The experiments on MAST summarised above relied on the injection of $100$ kW of RF power. Modelling showed that the majority of this power $(\sim95\%)$ was absorbed as EBW, which generate large values of $N_\perp$ $(N = k c/\omega)$, but can have values of $N_\parallel \approx 1$ \cite{Shevchenko_2010}. The electron density during the start-up phase of these experiments ranged from very small values to densities of $10^{18} \, \textrm{m}^{-3}$, and it is expected that the absorption will increase with the electron density until all power is absorbed, i.e. $\sigma \approx 1$.

\begin{figure}[!hbt]
	\centering
	\includegraphics[width=0.45\textwidth]{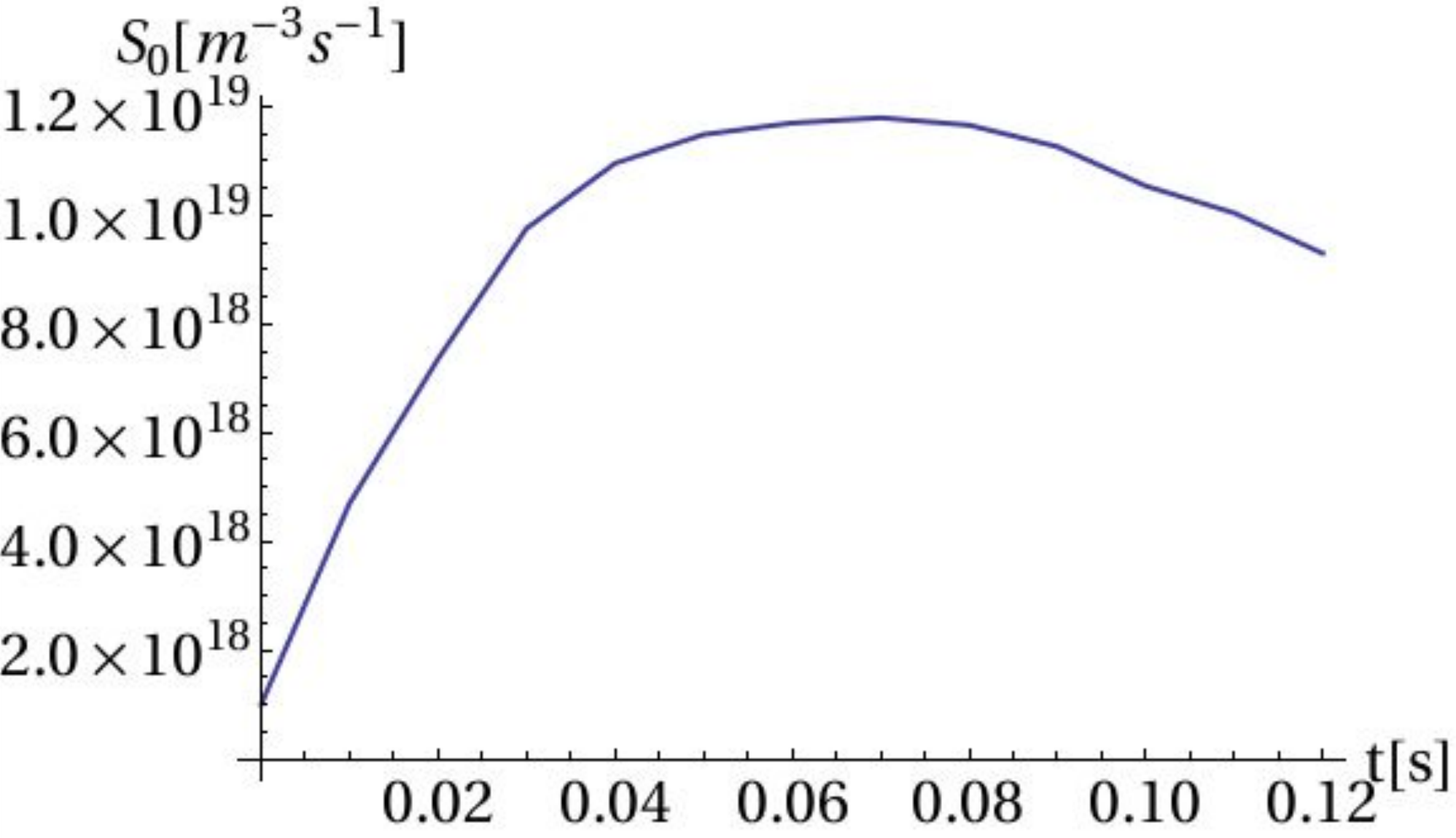}
	\caption[]{Simulated magnitude of the source term, $S_0$, which reproduces experimental densities.}
	\label{fig:source}
\end{figure}

\begin{figure}[!hbt]
	\centering
	\includegraphics[width=0.45\textwidth]{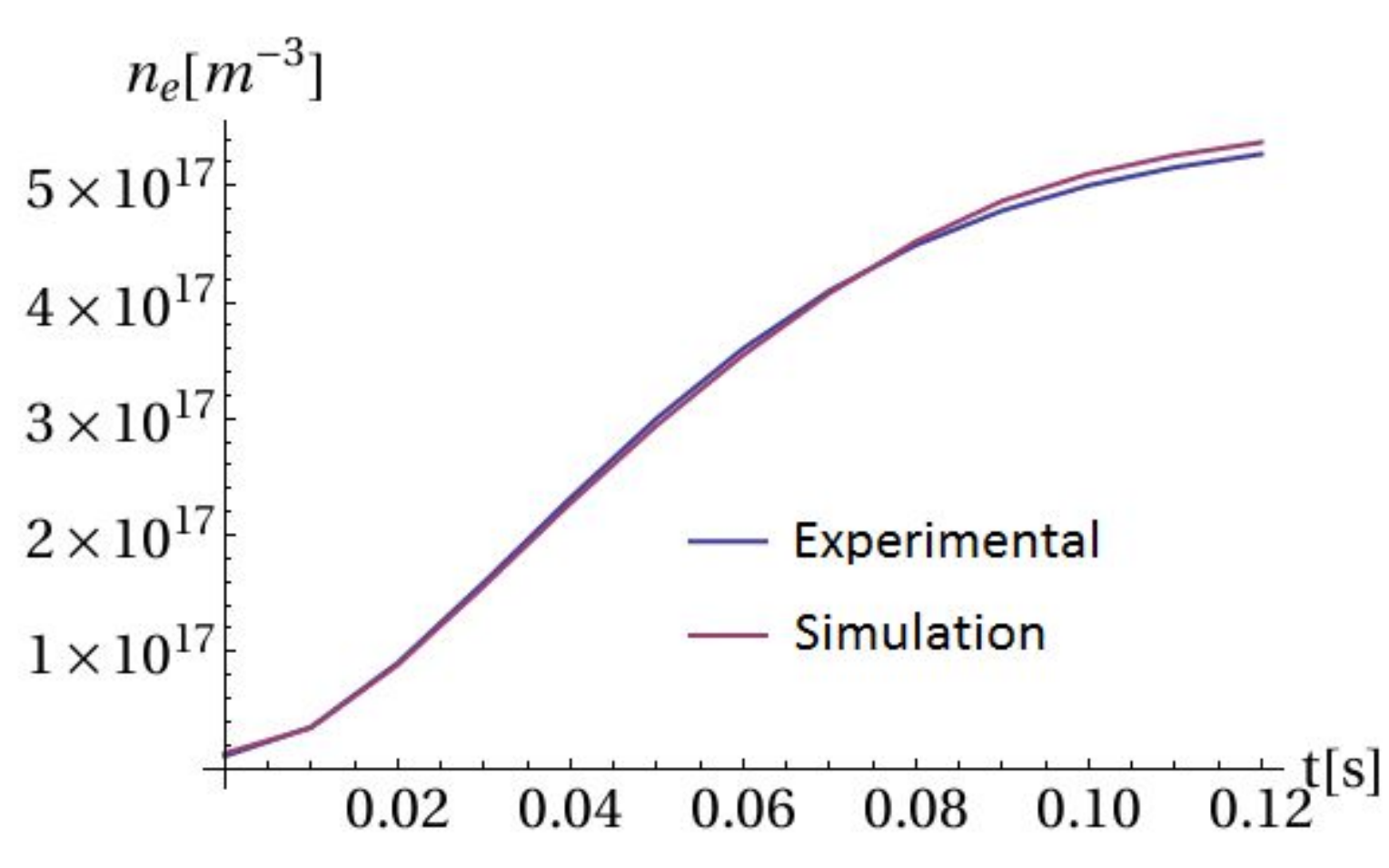}
	\caption[]{Experimental and simulated electron densities for EBW start-up.}
	\label{fig:density}
\end{figure}

\begin{figure}[!hbt]
	\centering
	\includegraphics[width=0.45\textwidth]{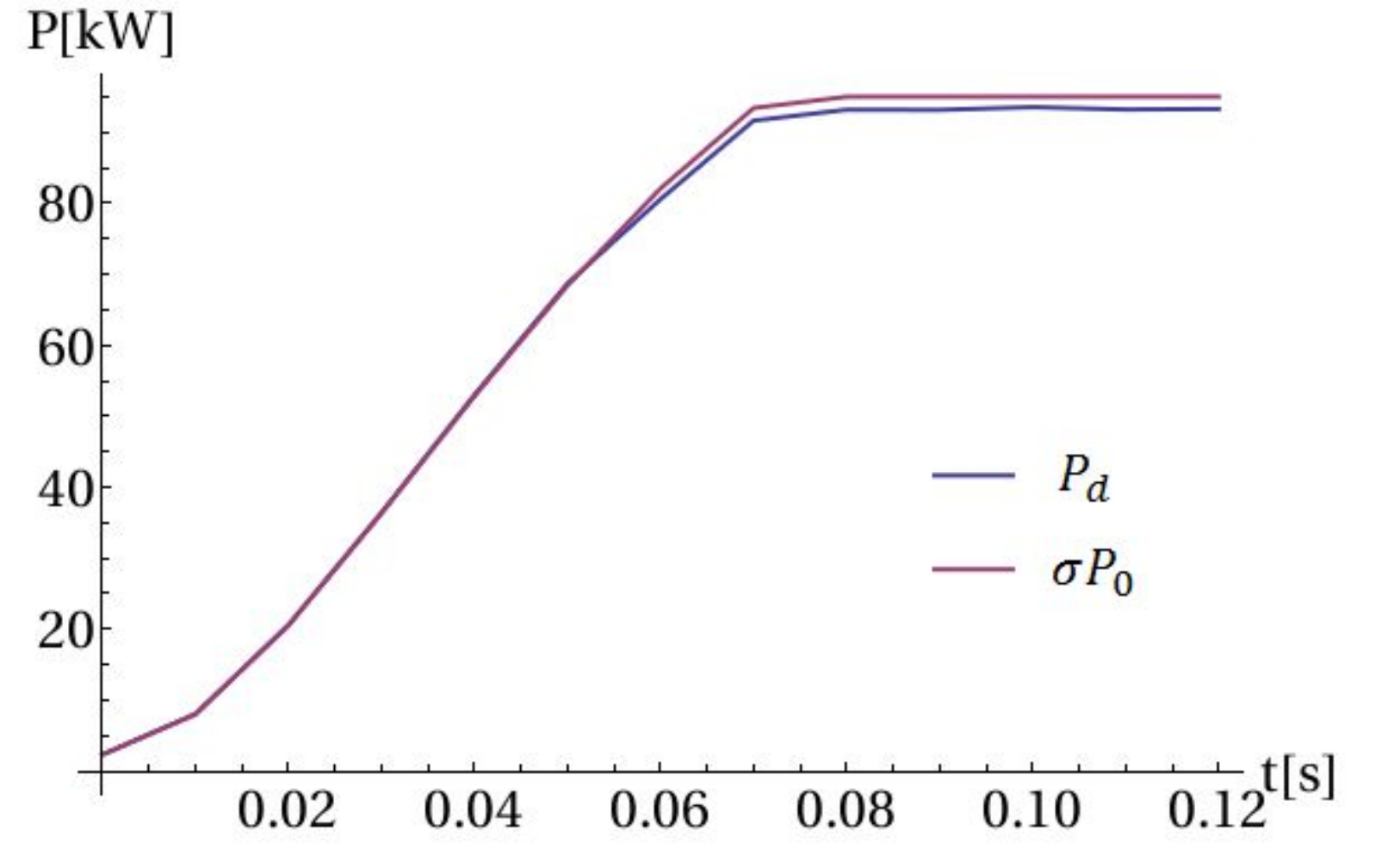}
	\caption[]{Simulated power absorbed for EBW start-up, calculated from the injected power and optical depth $(\sigma P_0)$ and from the distribution function $(P_d)$.}
	\label{fig:power}
\end{figure}

The simulation iterates over the source term, $S_0$ (fig. \ref{fig:source}) to reproduce experimentally observed densities (fig. \ref{fig:density}). At low densities the electron cyclotron resonance (ECR) and upper hybrid resonance (UHR) are close to each other, such that a small amount of power will be converted to EBW. As the density increases, however, the distance between the ECR and UHR increases and more power will be converted to EBW. The power converted to EBW is therefore assumed to be proportional to the electron density, while $\sigma = 1$. The absorbed power is shown in figure \ref{fig:power}.

\begin{figure}[!hbt]
	\centering
	\includegraphics[width=0.45\textwidth]{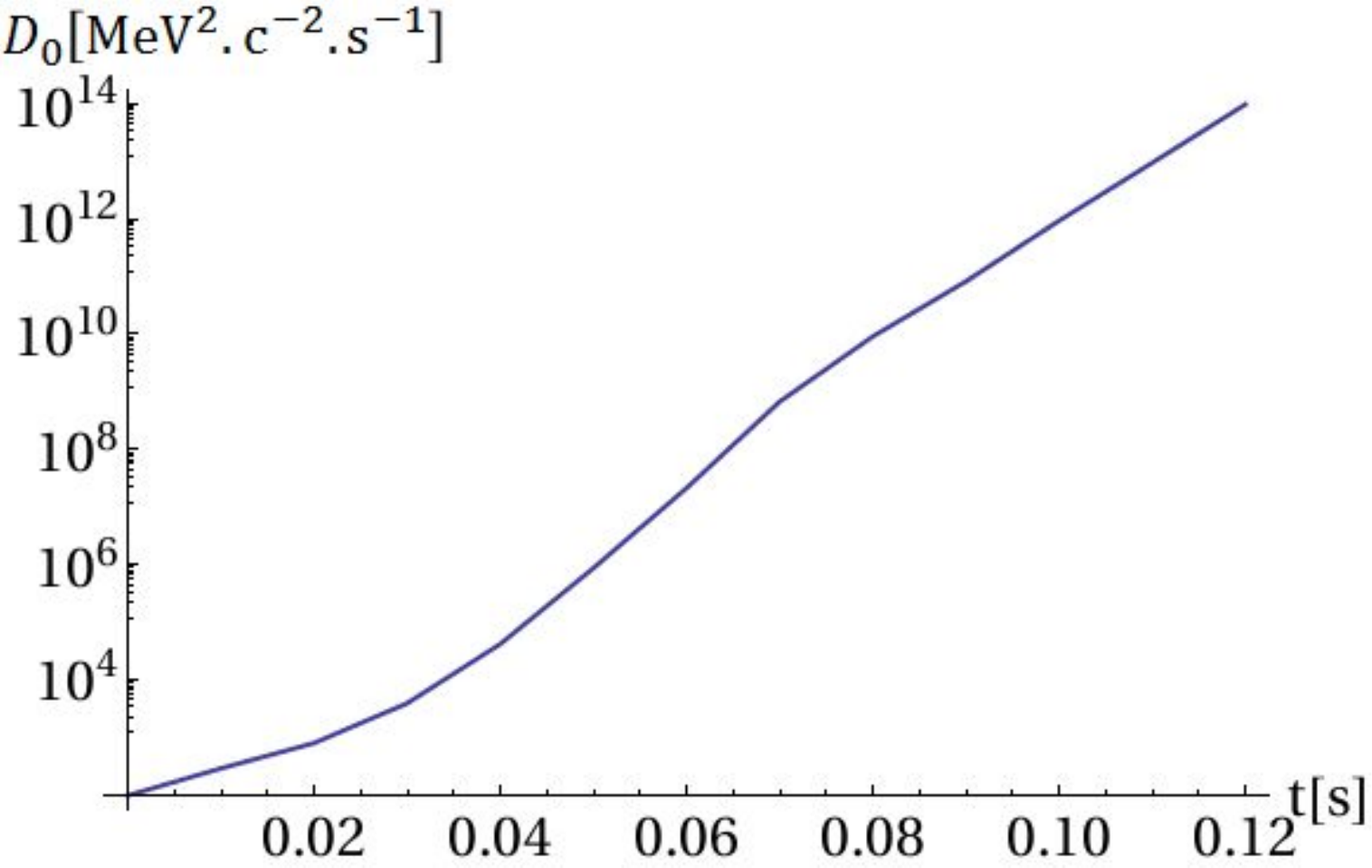}
	\caption[]{Simulated value for the diffusion coefficient $D_0$ for EBW start-up.}
	\label{fig:D0}
\end{figure}

The value of the diffusion coefficient $D_0$ can be determined such that the power absorbed calculated from the distribution function, $P_d$, equals the power absorbed found from the optical depth, $\sigma P_0$. The value of $D_0$, as a function of time, is shown in figure \ref{fig:D0}. A theoretical value for $D_0$ can be calculated, and such a value is a good initial value for $D_0$, but as the distribution function evolves in time and becomes distorted, the value of $D_0$ has to be updated to ensure the correct power is absorbed.

It is found that the distribution function flattens off, and therefore the value of $D_0$ keeps increasing to maintain full power absorption. At low power, the value of $D_0$ remains constant once full absorption is reached, as the distribution function remains close to Maxwellian. The effect of collisions might lessen the flattening of the distribution function at high power, and limit the growth of $D_0$.

The value of $D_0$ is updated through an iteration, under the criteria that the difference between $\sigma P_0$ and $P_d$ must be less than $5\%$. Typically, the value for $D_0$ that satisfies this criteria is found in no more than three iterations.

\begin{figure}[!hbt]
	\centering
	\includegraphics[width=0.45\textwidth]{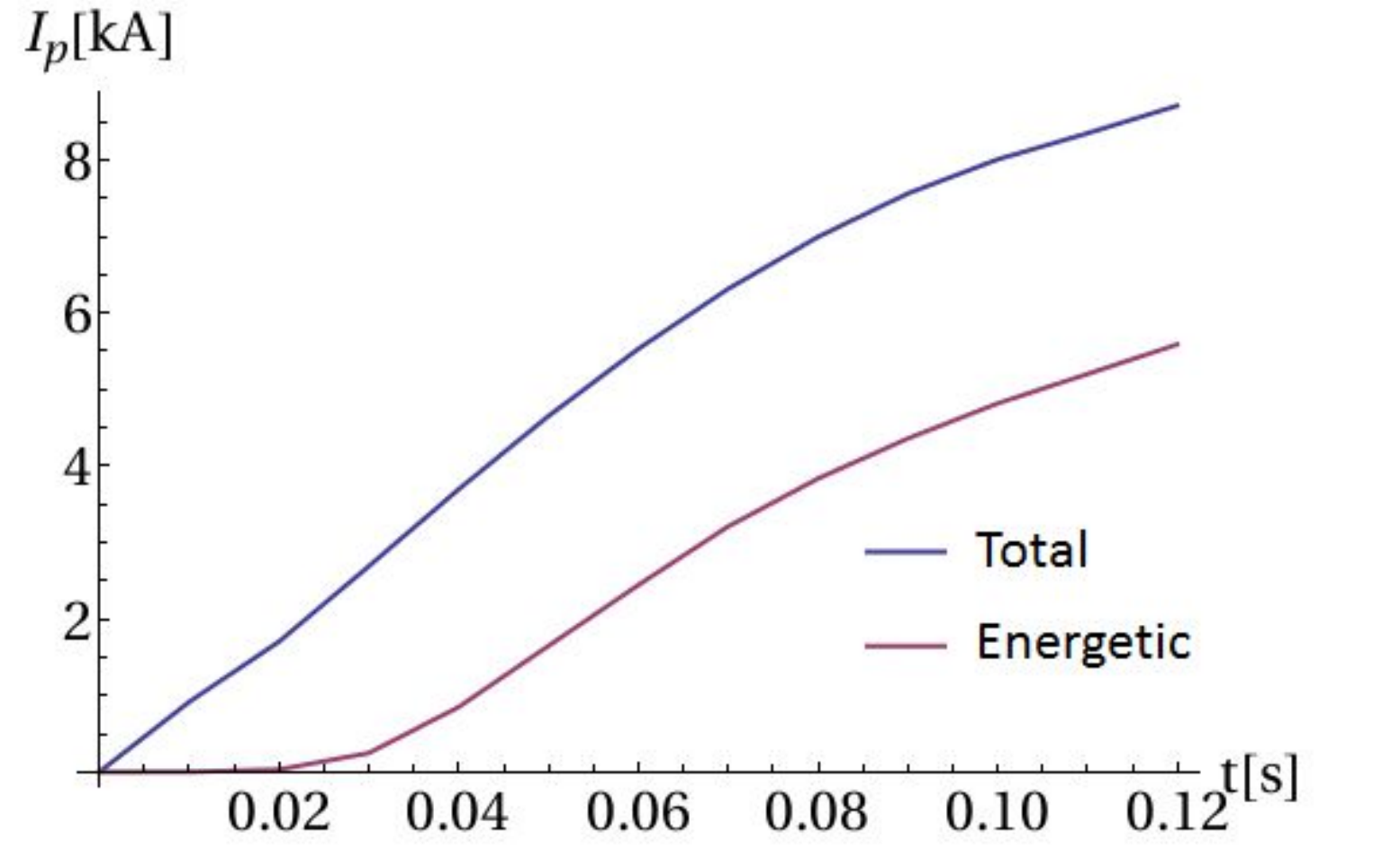}
	\caption[]{Simulated plasma current for EBW start-up. The total plasma current as well as the current carried only by energetic electrons is shown.}
	\label{fig:current}
\end{figure}

The calculated plasma current as a function of time is shown in figure \ref{fig:current}, with the fraction carried by energetic electrons $(T_e = 50-150 \, \textrm{keV})$ also shown. Experiments showed that the plasma current generated by EBW start-up is carried by supra thermal electrons with energies $T_e=70-100$ keV \cite{Shevchenko_2010, Shevchenko_2015}, and this is reproduced by simulations, albeit with a wider range of energies. 

\begin{figure}[!hbt]
	\centering
	\includegraphics[width=0.45\textwidth]{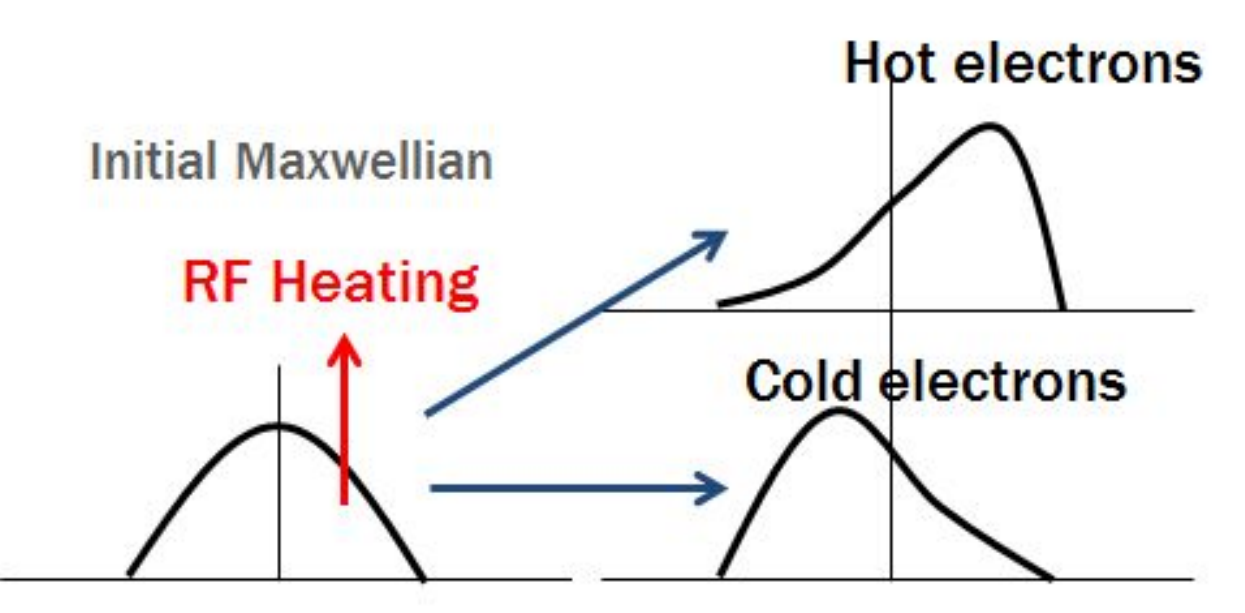}
	\caption[]{The preferential heating of electrons creates an asymmetry in both the hot and cold electrons, but in different directions.}
	\label{fig:mechanism}
\end{figure}

The preferential heating of electrons creates an asymmetry in both the hot and cold electrons, as shown in figure \ref{fig:mechanism}, which carries current in opposite directions. If the loss term were symmetric, this effect would not drive a net current, but due to the asymmetric loss term, electrons with negative $p_\parallel$ are lost faster than electrons with positive $p_\parallel$. This effect creates a larger population of electrons with positive $p_\parallel$, and as most of these electrons are energetic due to the interaction with EBWs, a net current is carried by hot electrons.

Although the generated current is nearly $10$ times smaller than the experimentally measured current, resulting in a current drive efficiency of $\sim 0.1$ A/W rather than $1$ A/W, this simple model does produce a population of supra thermal electrons in the same energy range as inferred in the experiments, with these electrons carrying a net current.

The discrepancy in the energy of electrons carrying the largest fraction of the plasma current can be attributed to the particular choice of parameters. The maximum energy attainable by electrons depend on the EBW heating term, and the specific wave parameters. As illustrated in figure \ref{fig:resonance}, there exists a maximum value of $p_\perp$ for which the diffusion term is non-zero, and this value determines the maximum energy attainable by electrons. Beyond this energy, the resonance condition can no longer be satisfied with real values of $p_\parallel$, and the electron can no longer interact with the EBW. A better choice of parameters would limit the energy attainable by these electrons to the same range as observed by experiments.

Inevitable uncertainties and approximations exist in this model, to the point that the choice of a different loss term, for example, may give better quantitative agreement with experiment. The inclusion of additional terms, such as collisions, might also play a role in the calculations, and will have to be included before quantitative comparisons between simulations and experiment can be made.

\section{Future Work}

Experiments have shown that CFS formed and that a tokamak like equilibrium was established during the EBW start-up phase. Further, a linear dependence between the plasma current and injected RF power was observed. In order to study these effects, the time evolution of the electron density has to be modelled. 

Existing, and more complicated, start-up models can be utilised to study the electron density. These models, such as the DYON code \cite{Kim_2012}, rely on the solution of power balance equations to obtain particle densities and temperatures.

For this paper, collisions have been neglected in order to study EBW start-up, as experiments suggest the plasma current is mainly carried by energetic electrons with energies above $70$ keV, and these electrons collide very infrequently. This observation was confirmed by simulations, but in order to study EBW start-up under a wider range of conditions, collisions will have to be included and studied.

Lastly, the EBW heating term relies on determining the power absorbed through the optical depth, and equating this to the power absorbed calculated with the distribution function. In this paper, it was assumed that the amount of power converted to EBW was proportional to the electron density, but a more sophisticated method would involve including a ray tracer to calculate the optical depth and conversion to EBW.

\section{Conclusion}
EBW current drive is a promising method for plasma start-up in spherical tokamaks, for which solenoid-free methods will be needed in burning plasma devices. A kinetic model for studying the time evolution of the electron distribution function is presented, from which the generation of plasma current can be calculated. Experimental evidence suggests that the current generated by EBW current drive is carried predominantly by supra thermal electrons, and this was confirmed by simulations, though the current drive efficiency was less than in experiments. The loss term is asymmetric during start-up, when the magnetic field lines have an open configuration, and this, coupled with a preferential heating of electrons, leads to an asymmetry in the distribution function. The EBW heating accelerates electrons to higher energies by increasing the perpendicular momentum, improving their confinement, to give a current carried by energetic electrons.

A method of relating the spatial dependent plasma wave interaction to a spatially independent diffusion term is also presented. This method relies on equating the volume averaged diffusion term to the absorption efficiency, obtained through knowing the optical depth and amount of power converted to EBW. The diffusion coefficient is chosen such that the power absorbed calculated from the distribution function will equal the power absorbed estimated from the amount of power converted to EBW.

Future studies will include investigating the effect of collisions, the solution of power balance equations to obtain the density evolution, as well as a ray tracer to obtain the wave parameters and absorption efficiency from an arbitrary distribution function.

\section*{Acknowledgements}
This work was part-funded by the RCUK Energy Programme under grant EP/I501045, the University of York, through the Department of Physics and the WW Smith Fund, and the Golden Key International Honour Society.

\end{document}